# Photoelectrocatalytic detection of NADH on n-type silicon semiconductors facilitated by carbon nanotube fibers

Melisa del Barrio[a,b]*, Moumita Rana[c], Juan José Vilatela[c], Encarnación Lorenzo[b,d], Antonio L. De Lacey[a], Marcos Pita[a]*

[a] *Instituto de Catálisis y Petroleoquímica, CSIC, C/ Marie Curie 2, 28049 Madrid, Spain*

[b] *Departamento de Química Analítica y Análisis Instrumental, Universidad Autónoma de Madrid, Ciudad Universitaria de Cantoblanco, 28049, Madrid, Spain*

[c] *IMDEA Materials Institute, C/ Eric Kandel 2, 28906, Getafe, Madrid, Spain*

[d] *IMDEA-Nanociencia, Ciudad Universitaria de Cantoblanco, 28049, Madrid, Spain*

Abstract

NADH is a key biomolecule involved in many biocatalytic processes as cofactor and its quantification can be correlated to specific enzymatic activity. Many efforts have been taken to obtain clean electrochemical signals related to NADH presence and lower its redox overpotential to avoid interferences. Suppression of background and secondary signals can be achieved by including a switchable electroactive surface, for instance, by using semiconductors able to harvest light energy and drive the excited electrons only when irradiated. Here we present the combination of a n-type Si semiconductor with fibers made of carbon nanotubes as electroactive surface for NADH quantification at low potentials only upon irradiation. The resulting photoelectrode responded linearly to NADH concentrations from 50 μM to 1.6 mM with high sensitivity (54 μA cm$^{-2}$ mM$^{-1}$). This system may serve as a biosensing platform for detection and quantification of dehydrogenases' activity.

1. Introduction

Nicotinamide adenine dinucleotide (NAD) and its reduced form (NADH) are an important redox couple found in all living systems. They act as enzymatic cofactors and play a key role in biological electron transport. NADH is involved in the reactions catalysed by more than 300 dehydrogenase enzymes [1], in ATP synthesis, and also other non-redox processes such as repairing the DNA system [2,3]. Moreover, altered NADH levels are linked to various neurodegenerative diseases [4–7] and mental disorders [4,8,9]. Given its biochemical and medical interest, several methods have been employed to detect NADH, including high performance liquid chromatography [10], fluorescence [11,12], electrochemiluminesce [13–15], electrochemical [3,16,17], and photoelectrochemical [18–24] methods.

In the last decades, research efforts have been focused on the electrochemical oxidation of NADH to $NAD^+$ [25,26] due to the advantages of electrochemical detection, such as simple instrumentation and high sensitivity. Although the formal redox potential of the $NAD^+$/NADH couple is -0.560 V, NADH shows slow electron transfer kinetics at unmodified electrodes and overpotentials larger than 1 V need to be applied for the oxidation of the analyte, which increases the risk of interferences. Fouling of the electrode surface due to the adsorption of oxidation products, which leads to electrode passivation and therefore irreproducibility of the analytical signal, has also been observed [1,27]. To overcome the problems of bare electrodes, the modification of the electrode surface with redox mediators (e.g. quinone derivatives, phenazine, phenoxazine and phenothiazine dyes, nitro-fluorenone and catechol) which catalyze the oxidation of NADH [17,28,29], polymers [3], and nanomaterials, such as carbon nanotubes and graphene sheets [16], have been widely considered. Physical adsorption, electropolymerized films, and to a lesser extent, covalent attachment have been used for the electrode modification with redox mediators to obtain a stable immobilization. Carbon nanotubes show electrocatalytic activity towards NADH oxidation and minimize surface fouling, but these carbon nanomaterials have low solubility and, consequently, involve tedious protocols for electrode modification. Preliminary results on electrocatalytic NADH oxidation on carbon nanotube fiber microelectrodes were reported by Wang *et al.* [30], and only some optimizations have been proposed since then by the use of redox mediators and specific pretreatments [31].

More recently, photoelectrochemical (PEC) detection methods have become very popular due to their attractive advantages. They inherit the benefits of electrochemical detection and represent a highly sensitive and low background signal approach, as they use a unique signal transducing modality involving separate processes for signal generation (photoexcitation) and detection (electrochemical) [32]. PEC biosensing devices have been applied in nucleic acid analysis, immunoassays, cell detection, protein/enzyme sensing, and the monitoring of small molecules [33–36], such as those that are produced during enzymatic reactions (e.g. oxygen, hydrogen peroxide, or NADH). Semiconductors are key components of PEC systems, as they allow the switching ON and OFF upon illumination [33,35]. Most PEC sensors found in the literature are based on quantum dots (QDs) and $TiO_2$ [37], often combined with other photoactive materials to improve their PEC behaviour. For instance, traditional QDs (e.g. CdSe and CdS) exhibit low photon-to-current conversion efficiencies, due to their relatively low quantum yields, and are prone to photobleaching, which results in unstable photocurrents. Therefore, QDs-based PEC platforms include other semiconducting materials, like graphitic carbon nitride or metallic oxides [33]. Furthermore, a new class of QDs, CsPbBr$_3$ perovskite QDs, exhibited better PEC characteristics and was used, in combination with ZnO inverse opal photonic crystals, for NADH detection [18]. Nevertheless, the toxicity of the metal ions from QDs (Cd, Se, Pb…) and the poor stability of perovskite are still an issue. Regarding TiO$_2$ nanoparticles, apart from being subject to some of the problems previously indicated for QDs, they can only use UV light for due to the inherent band gap of this semiconductor, which limits their applications. The construction of hetero-structured TiO$_2$ photoelectrodes by coupling this material with another semiconductor having a smaller band gap or a dye sensitizer have been proposed to overcome this issue [37]. The generation of a Ti-dopamine charge transfer complex was used for the design of the only TiO$_2$ photoelectrode reported for PEC sensing of NADH under visible irradiation [23].

Here we show the photocatalytic detection of NADH at low potential on a simple platform composed of a n-type silicon semiconductor and a carbon nanotube fiber film (Si/CNT$_f$). Si has a small band gap of 1.1 eV, which enables visible-light-driven oxidation reactions. Carbon nanotube fibers were easily deposited on the semiconductor surface without further modifications with any redox mediator. In spite of the unique properties of this carbon nanomaterial, such as excellent conductivity, high surface area and stability [38,39], they have barely been exploited for NADH determination. First, we characterize both CNT$_f$ and Si/CNT$_f$ photoelectrodes by scanning and transmission electron microscopy, Raman and impedance spectroscopy. Then, we show the photocatalytic oxidation of NADH on Si/CNT$_f$ photoelectrodes and we investigate the origin of the photooxidation signal. Furthermore, the analytical performance and stability of the photoelectrodes are evaluated.

## 2. Experimental

### 2.1. Reagents

Sodium hydrogen phosphate, sodium dihydrogen phosphate dihydrate, β-nicotinamide adenine dinucleotide reduced disodium salt hydrate (NADH), n-type silicon <111> wafers doped with phosphorous, fluorine-doped tin oxide (FTO) coated glasses, potassium ferricyanide, potassium ferrocyanide trihydrate, ferrocene, thiophene, and toluene were purchased from Merck. 3-Aminopropyltriethoxysilane 99% (APTES) and toluene 99.9% were obtained from Fisher Scientific. Grey dielectric paste was supplied by Gwent Group. Fresh MilliQ-grade water purified with Millipore filtering system water (18.2 Ω·cm) was used to prepare aqueous solutions.

### 2.2. Instruments

Photoelectrochemical measurements were performed in a three-electrode cell configuration using a Pt wire as a counter electrode and an Ag/AgCl (3 M NaCl) reference electrode. The cell was filled with 5 mL of 100 mM phosphate buffer pH 7.4. The measurements were carried out at room temperature and controlled with a microAutolab II potentiostat from Metrohm. Illumination was carried out with a LOT Quantum Design 150 W Xenon lamp with the output power set to 81 W, so that the power reaching to the electrode surface corresponded approximately to one average Sun intensity. A 1 mm-thick polycarbonate UV filter was used to remove the radiation below ≈380 nm. Electrochemical impedance spectroscopy (EIS) experiments were performed in an equimolar $[Fe(CN)_6]^{4-}/[Fe(CN)_6]^{3-}$ solution (0.5 mM each in phosphate buffer) at 0.2 V *vs* Ag/AgCl by applying an alternating current (AC) amplitude of 10 mV in a frequency range from $10^4$ to 0.1 Hz, using an Autolab PGSTAT30/FRA2 potentiostat/galvanostat from Metrohm. EIS data was fitted using electrical equivalent circuits with ZView2 software from Scribner Associates Inc. (USA). The microscopic structure of the CNT fibers were investigated using field emission scanning electron microscopy (FESEM, Helios Nanolab 600i FIB-FEGSEM dual-beam microscope, at 5 kV) and transmission electron microscopy (TEM, Talos F200X FEG, 80 kV). Thermogravimetric analysis (TGA) was performed using a Q50 thermobalance from TA Instruments under aerial condition. Wetting was analysed by determining spreading rate and equivalent contact angle through image analysis of liquid drops on porous CNT fibre samples, as described previously [40].

2.3. Preparation of carbon nanotube fibers ($CNT_f$)

To synthesize the CNT fibers, a floating catalyst chemical phase vapor deposition (FC-CVD) process was used where toluene, ferrocene and thiophene as carbon, catalysts and promoter sources, respectively. The precursors were injected continuously from the top of a vertical tube furnace maintaining a temperature of 1250 °C under $H_2$ atmosphere to produce aerogel of long CNTs floating in the gas phase. The continuous fibers of CNT were mechanically withdrawn at the bottom of the tube furnace and winded on a rotating spool leading to a unidirectional non-woven fabric.

2.4. Preparation of the $Si/CNT_f$ photoelectrodes

The procedures for cleaning and silanizing the silicon wafers were adapted from a previous work using a different silane coupling agent [41]. Silicon wafers were cut into 0.8 x 2 cm chips and cleaned by immersing them in a solution composed of 1:1:1 ammonium hydroxide (30%), hydrogen peroxide (30%) and water at 60ºC for 1 h. Then the chips were rinsed with water and dried with an Ar stream. The surface was then modified with a 2% APTES solution in dry toluene for 15 min at 60 ºC. The silanized surface was sequentially rinsed with toluene, ethanol and water and dried in an oven at 110 ºC. The $CNT_f$ were deposited as reported before [42]. First, the $CNT_f$, covered on both sides by protective plastic sheets, were cut into 0.5 x 0.5 cm pieces using scissors. After removing the plastic sheet, one piece was placed on the surface of Si electrodes with the help of PTFE tweezers. Finally, a drop of acetone was added, wetting the $CNT_f$, while the edges of the piece were held to prevent the piece getting folded or rolled up. A flat and well attached layer was obtained. Afterwards, the electrodes were heated at 150 ºC. This thermal treatment, together with the Si modification with the silanization agent (APTES) improved the charge transfer and the stability of the photoresponse. Finally, the exposed Si surface of the electrode was completely masked with PTFE tape, except for a small area (width = 1 mm) around the $CNT_f$.

2.4. Preparation of FTO/$CNT_f$ photoelectrodes

FTO-covered glasses were cut into 2 x 3 cm slides and cleaned as reported in [43], by sequentially immersing them in water, ethanol and acetone in an ultrasonic cleaner for 15 min. Afterward, the slides were let to dry. A copper wire was fixed to one edge of the FTO surface with conductive adhesive tape. A 1 cm diameter piece of $CNT_f$ was then placed on the FTO surface by following the procedure described in the previous section for Si electrodes. The FTO electrode was then masked with dielectric paste.

3. Results and discussion

3.1. Characterization of the photoelectrodes

Due to their inherent hydrophobic nature, the as-prepared $CNT_f$ fabric can hardly interact with the aqueous electrolytes. Therefore, to improve their hydrophilicity the pristine fibers were exposed to ozone, generated in situ from a UV source under aerial atmosphere for 30 minutes each side [44]. This helps in oxidizing the CNT surface, as observed by the Raman spectroscopy. Figure 1a shows the comparison of the Raman spectra of the pristine $CNT_f$ with the ones treated with ozone. The peaks around 1355, 1586 and 2691 cm$^{-1}$ can be attributed to the symmetric vibrations from the defected graphite network (D band), the crystalline graphitic regime (G band) and overtone of the D band (2D band) of carbon nanotube, respectively [40]. Upon functionalization, the intensity ratio of the D to G peak increases from 0.2 (± 0.02) to 0.5 (± 0.04). This establishes the presence of newly evolved oxygen functional groups (e.g. C=O, C-O, O-C=O) upon ozone treatment [44]. By comparing the spread/wicking rate of water in the pristine and functionalised $CNT_f$ samples we could directly observe the transition to a hydrophilic material (Figure S1), and calculate the equivalent contact angle of water using the following equation [40,45]:

$$\frac{S}{S_0}\left[ln\frac{S}{S_0} - 1\right] = -1 + \frac{2\pi r \gamma cos\theta}{3\eta S_0} t \qquad (1)$$

Here S and $S_0$ are the wicking area at time t and 0, respectively, r is the equivalent capillary radius, η the viscosity, γ is the surface tension of the liquid and θ is the equivalent contact angle (for a smooth, flat surface in equilibrium). As shown in Figure S1, the 2 μL droplet remains immobile on the pristine $CNT_f$ fabric, whereas it spreads fast on the functionalized sample, over a few seconds. From the plot of $(S/S_0)(ln(S/S_0)-1)$ against time (Figure S1f), we estimated the contact angle of water on the functionalized simple to be 67.2°, which confirms the increased hydrophilicity of the functionalized sample.

Figure 1b-c show the FESEM images of the CNT fibers. As shown in the figure 1c, the ultra-long CNT fibers entangle with each other to form the macroscopic fabric. TEM investigations reveals that CNT fibers are a porous network of interconnected

CNT bundles with diameter in the range of 8-21 nm (Figure 1d, e). CNT content and residual catalyst produced during the FC-CVD process can be extracted from thermogravimetric analysis of the CNT fibers (Figure S2).

The $CNT_f$ were deposited on the Si semiconductor surface, which was previously modified with a silanization agent (APTES), and the resulting $Si/CNT_f$ were thermally treated as described in the experimental section. We used electrochemical impedance spectroscopy to analyze the charge transfer at the $Si/CNT_f$ electrodes, using $[Fe(CN)_6]^{4-}/[Fe(CN)_6]^{3-}$ as a redox probe. The charge transfer resistance, $R_{ct}$, is equal to the diameter of the semicircle extrapolated in the Nyquist plot, in which the real part of the impedance (Z') is plotted along the x-axis and the imaginary part (Z'') is plotted on the y-axis (each data point is the impedance at one frequency, with frequency decreasing from left to right). [46] Figure 2A shows the Nyquist plots of $Si/CNT_f$ and silanized Si. The $R_{ct}$ values obtained after fitting to their equivalent circuits (see Figure 2B) are collected in Table 1. In the absence of irradiation, high impedance values were observed for $Si/CNT_f$ (grey squares in the figure) and a large $R_{ct}$ of 303 kΩ cm$^{-2}$ was found. A 20-fold decrease in the impedance at 1 Hz (lowest frequency applied with corresponds with the far-right data point of the spectra) and a 40-fold reduction in the charge transfer resistance (see Table 1) was achieved upon light irradiation of the $Si/CNT_f$ photoelectrode (red squares). The silanized Si semiconductor (red triangles in Fig. 2A) exhibited a large charge transfer resistance ($R_{ct1}$ in the equivalent circuit indicated in Fig. 2B = 356 kΩ cm$^{-2}$) upon light. The smaller semicircle shown in the high frequency region of the Nyquist plot (left side of the plot) can be assigned to lower charge transfer resistance ($R_{ct2}$ in the equivalent circuit indicated in Fig. 2B) of the negatively charged redox probe at the positively charged APTES/electrolyte interface. A bare (non-silanized) Si electrode was also analyzed. Figure S3A shows that the silanization greatly decreased the charge transfer resistance of the semiconductor. Moreover, the $R_{ct}$ of Si/CNT$_f$ fell to half the value of that of a Si/CNT$_f$ photoelectrode which was not thermally treated (Fig. S3B). Therefore, the thermal treatment applied to the photoelectrode, together with silanization, were a successful procedure to enhance the charge transfer.

Table 1. Charge transfer resistance with (ON) and without (OFF) irradiation obtained by fitting the Nyquist plots to the equivalent circuits shown in Figure 2.

|  | $R_{ct}$ (k$\Omega$ cm$^{-2}$) |
| --- | --- |
| Si/CNT$_f$ OFF | 303±4 |
| Si/CNT$_f$ ON | 7.79±0.08 |
| Silanized Si ON | 356±4 |

3.2. Photoelectrochemical detection of NADH

We investigated the capability of Si/CNT$_f$ electrodes to photooxidize NADH by irradiating the frontal side of the electrode, showing the CNT$_f$ layer. Figure 3 shows the cyclic voltammograms in the absence and in the presence of irradiation. Upon illumination, a decrease of 0.35 V in the overpotential of NADH oxidation and a 5-fold increase in the current value at about 0.3 V, compared to the measurement without irradiation, were observed. Thus, irradiation significantly reduced the overpotential for NADH oxidation on Si/CNT$_f$ electrodes, without the need of any conventional redox mediator.

We also used linear sweep voltammetry to examine the photoelectrochemical response upon chopped irradiation with and without NADH. The upper limit of the applied potential was 0.4 V *vs* Ag/AgCl, since more oxidant potentials affected the stability of the photoresponse (Fig. S4). In particular, we observed a gradual increase in the onset potential of NADH photooxidation after voltammetry experiments at higher potential limits. When we swept up to 0.4 V, the overpotential initially increased and after a few scans, the oxidation peak stabilized at around 0.3 V. The shift in the NADH oxidation overpotential correlated with higher charge transfer resistances measured by electrochemical impedance spectroscopy experiments (Fig. S5). The increase of the photocurrent in the presence of the NADH was observed at a potential as low as -0.1 V (Figure 4), while lower currents were obtained in the absence of NADH. The silanized Si electrode exhibited only negligible photocurrents both in phosphate buffer (data not shown) and in the presence of NADH (green trace in Figure 4). This can be explained by the relatively low conductivity of the Si wafers used ($10^{-3}$-40 Ω cm), even upon irradiation. Therefore, the presence of $CNT_f$ on the electrode was crucial to detect NADH photooxidation. Herein, the edge plane and functional groups in the CNTf are responsible for the adsorption and catalytic oxidation of NADH [47,48], and the extended conducting network of the $CNT_f$ assists in efficient charge transfer, thereby lowering the overpotential of the NADH oxidation.

To investigate the origin of the NADH photooxidation response, we performed measurements upon back irradiation of the masked (in which the back surface of silicon was covered by PTFE tape) and unmasked Si/$CNT_f$ electrode. When the unmasked electrode was illuminated in this configuration, the cyclic voltammograms remained almost the same as those with frontal irradiation shown above (Fig. S6), whereas for the masked electrode a lower photocurrent was observed and the NADH oxidation peak did not appear at potential lower than 0.4 V (Fig. S7). These results suggest that: (1) the photoelectrocatalytic NADH activity is not caused by photoexcitation of the $CNT_f$ because it occurred even though $CNT_f$ were not directly irradiated by the back illumination; (2) the photoexcitation of Si is indeed required because NADH oxidation at lower overpotential was not measured with the masked electrode; (3) the photoexcitation of Si is not the rate limiting step of the overall photoelectrocatalytic process, as the photocurrent of the unmasked electrode showing a larger Si surface exposed to back irradiation ($\approx 0.8$ cm$^2$) and the masked electrode covered by the CNT$_f$ and PTFE tape (except from a $\approx 0.1$ cm$^2$ around the CNT$_f$) upon frontal irradiation were almost equal. As the transmittance of CNT$_f$ is almost 0% (Fig. S8), the Si surface covered by them was not exposed to light and therefore we discarded a hypothetical photoexcitation of Si beneath CNT$_f$.

To further investigate the role of CNT$_f$ on the NADH photooxidation, we studied the photoresponse of a FTO-based electrode (FTO/CNT$_f$), which possesses a lower resistivity than that of Si, by linear sweep voltammetry with chopped light. The FTO/CNT$_f$ electrodes showed a NADH oxidation wave at 0.53 V and a greater oxidation current than the bare FTO electrode, but we did not observe any enhancement in the current density upon irradiation (Fig. S9). Furthermore, we showed above that R$_{ct}$ for Si decreased after CNT$_f$ deposition, which means that CNT$_f$ increase the charge transfer at the electrode. Consequently, the photoelectrochemical response of Si/ CNT$_f$ electrodes can only be due to the photoexcitation of Si and the efficient charge transfer provided by the CNT$_f$.

On the basis of such notable photocurrent values achieved at low overpotential with Si/CNT$_f$ electrodes, we studied the influence of NADH concentration on the photoresponse. Since the NADH oxidation photocurrent increased gradually with the applied potential up to ca. 0.30 V, photoelectrochemical experiments were performed at 0.2 V to improve the signal-to-noise ratio, while minimizing the effect of possible interferences. Figure 5 shows the results of the experiments in a range of NADH concentrations between 50 μM and 1.6 mM. Two repeated irradiation cycles were performed for each concentration. The photocurrent density (j) linearly increased with NADH concentration in the range measured according to the following equation: j (µA cm$^{-2}$) = 53.7 [NADH] (mM) + 0.566. The fit yielded a R² = 0.998, for 3 independent electrodes. A high sensitivity of 54±12 µA cm$^{-2}$ mM$^{-1}$ was obtained. The detection limit was calculated to be 1±0.7 µM. The Si/CNT$_f$ electrodes showed a broader linear range than most photoelectrochemical approaches reported so far for NADH oxidation and one of the highest sensitivities (Table 2) [18,20–24]. We have found one system in the literature that incorporates a n-type silicon semiconductor as electrode substrate [20], although it does not act as light absorbing material as in our Si/CNT$_f$ system does. After the irradiation cycle and photoelectrochemical NADH oxidation, the initial current (in the dark) is recovered fast and the electrode can be further used. It should be noted that the ON-OFF photo-chronoamperometric experiments yielded a capacitive discharge when switching the light on. This process is attributed to the absorption of counter-ions when the light is off that are released upon illumination and its concomitant conduction increase. Very small relative standard deviations (<2% for all tests performed) were found for 2 irradiation cycles on the same electrode. The stability of the electrode was further studied for a larger number of repeated irradiation cycles (Figure 6). 95% of the initial photocurrent was retained after 25 cycles, which proved excellent repeatability. Such electrode stability can be attributed to the reversible redox nature of the catalytic sites in the CNT fiber as well as absence of electrode passivation from the strong adsorption of the NAD$^+$ produced during NADH oxidation, which is very common for other carbon-based electrodes [49–51]. The high sensitivity achieved also seems to support the absence of such passivation and parasitic reactions. This was also aided by the low potential applied for NADH detection that excludes typical interferences appearing at higher potential values.

Table 2. Performance comparison of PEC approaches for NADH determination

| Photoelectrode | Sensitivity (µA cm$^{-2}$ mM$^{-1}$) | Linear range (µM) | Refs. |
| --- | --- | --- | --- |
| FTO/ZnO/CsPbBr$_3$ | 15 | 0.1 - 250 | [18] |
| Si/InGaN/GaN nanowires | 91 | 5 - 50 | [20] |
| FTO/Polydopamine–Nile Blue/Fe$_2$O$_3$ | 5 | 100 - 2000 | [21] |
| Glassy carbon/CuWO$_4$ nanoplates | 140 µA mM$^{-1}$ [a] | 10 - 50 | [22] |
| ITO/dopamine-TiO$_2$ | 2 | 0.5 - 120 | [23] |
| Glassy carbon/raspberry shaped ZnO-Au nanostructures | 5 | 1 - 7 | [24] |
| Si/CNT$_f$ | 54 | 50 - 1600 | This work |

[a] unknown geometric area of the electrode

4. Conclusions

In this study, we have shown visible-light-triggered detection of NADH using a n-type silicon semiconductor and carbon nanotube fibers. We have demonstrated that the NADH photooxidation signal is generated by the photoexcitation of silicon and the effective charge transfer of CNT$_f$. The presence of CNT$_f$ is crucial to provide the catalytic sites for NADH oxidation and, to the best of our knowledge, this is the first time they have been applied in a photoelectrocatalytic system for NADH detection. The Si/CNT$_f$ photoelectrodes allowed NADH determination at low potential (0.2 V *vs* Ag/AgCl) in a broad linear range of 50 µM -1 mM with high sensitivity. Moreover, the photoelectrodes displayed excellent stability for repetitive NADH detection, which confirms the reversible nature of the CNT$_f$ redox sites and the absence of parasitic reactions to produce catalytic poisons fouling the surface. These results contribute to the development of light-switchable NADH sensors or biosensors by combination with NADH-dependent enzymes.

## Acknowledgements

M.B. acknowledges funding from the European Union's Horizon 2020 research and innovation programme under the Marie Skłodowska-Curie grant agreement No. 713366. A.D.L., M.P and J.J.V. thank the "Comunidad de Madrid" for its support to the FotoArt-CM project (S2018/NMT-4367) through the Program of R&D activities between research groups in Technologies 2018, co-financed by European Structural Funds. J.J.V. is grateful for generous financial support provided by the European Union Horizon 2020 Program under grant agreement 678565 (ERC-STEM) and by the MINECO (RyC-2014-15115, HYNANOSC RTI2018-099504-A-C22).

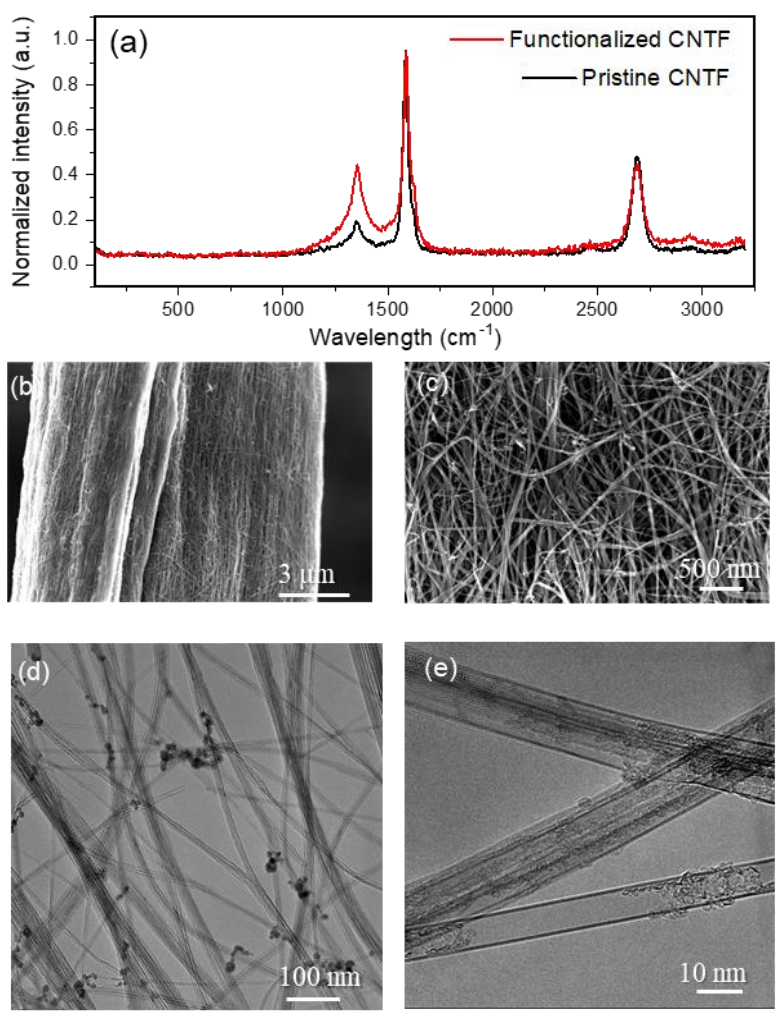

**Figure 1.** Characterizations of CNT fibers. (a) Raman spectra of the pristine and ozone treated CNT fibers. (b, c) FESEM and (d, e) TEM images of the CNT fibers.

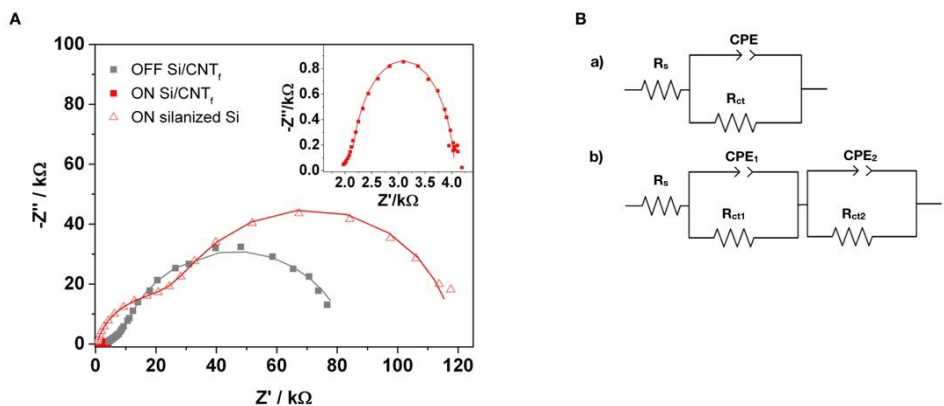

**Figure 2.** (A) Nyquist plot curves of Si/CNT$_f$ upon irradiation (red squares, ON) and without irradiation (grey squares, OFF) and a silanized Si semiconductor upon irradiation (red triangles). The lines represent the fit to the equivalent circuits indicated in panel B. Inset in panel A: Magnification of the Nyquist plot of Si/CNT$_f$ upon irradiation. (B) Equivalent circuit of a) Si/CNT$_f$ and b) silanized Si. Conditions: 1 mM [Fe(CN)$_6$]$^{4-}$/[Fe(CN)$_6$]$^{3-}$ solution in 0.1 M phosphate buffer pH 7.5, 0.2 V *vs* Ag/AgCl, $10^4$-1 Hz ($10^4$-0.1 Hz for the Si semiconductor), 170 mW cm$^{-2}$.

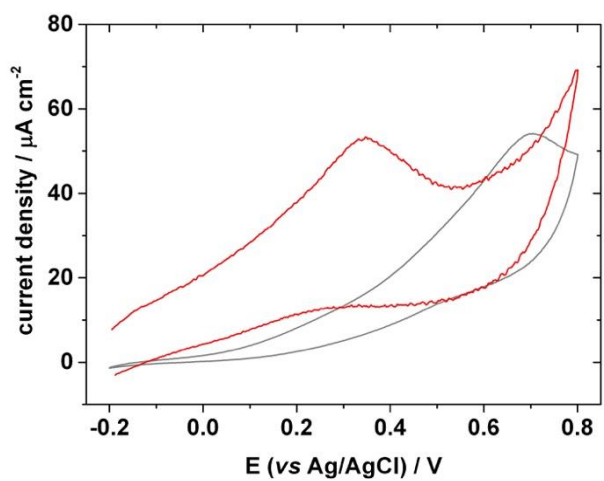

**Figure 3.** Cyclic voltammograms of Si/CNT$_f$ upon irradiation (red) and without irradiation (grey) in the presence of 1 mM NADH. Conditions: 0.1 M phosphate buffer pH 7.5, scan rate = 20 mV s$^{-1}$, 170 mW cm$^{-2}$.

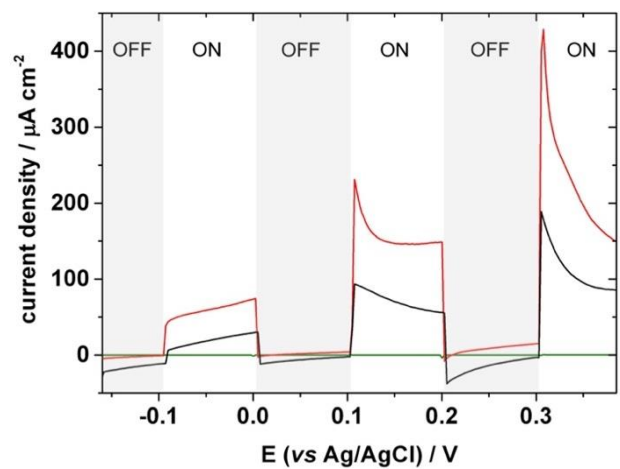

**Figure 4:** Linear sweep voltammograms of the Si/CNT$_f$ electrode upon chopped irradiation in the presence (red) and absence of 1 mM NADH (black), and of a silanized Si electrode in the presence of 1 mM NADH (green). Conditions: 0.1 M phosphate buffer pH 7.5, scan rate = 20 mV s$^{-1}$, 170 mW cm$^{-2}$.

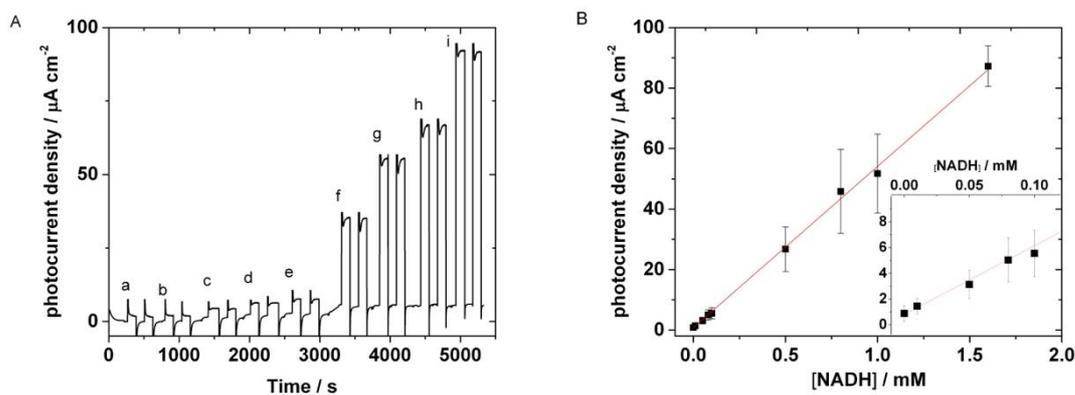

**Figure 5:** (A) Photocurrent response of Si/CNT$_f$ to increasing concentrations of NADH (a: blank, b: 0.010 mM, c: 0.050 mM, d: 0.080 mM, e: 0.100 mM, f: 0.500 mM, g: 0.800 mM, h: 1 mM, i: 1.6 mM). (B) Calibration plot (n=3 electrodes) (photocurrent density (μA cm$^{-2}$) = 53.7 [NADH] (mM) + 0.566; R² = 0.998). Conditions: 0.1 M phosphate buffer pH 7.5, E=0.2 V *vs* Ag/AgCl, 170 mW cm$^{-2}$.

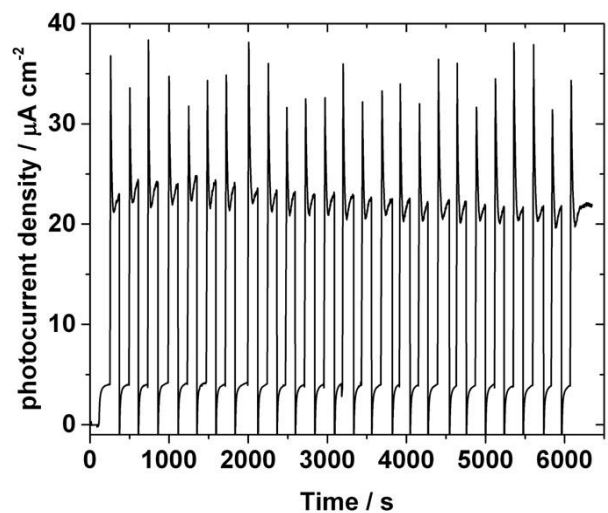

**Figure 6:** Stability of the photocurrent response of the Si(APTES)/CNT$_f$ electrode for 25 consecutive measurements (2 min irradiation cycles) in the presence of 0.5 mM NADH. Conditions: 0.1 M phosphate buffer pH 7.5, E=0.2 V *vs* Ag/AgCl, 170 mW cm$^{-2}$.